\documentclass[twocolumn,amsmath,amssymb,aps,prl,groupedaddress,floatfix,showpacs, superscriptaddress]{revtex4-1}
\usepackage[utf8]{inputenc}
\usepackage{graphicx}% Include figure files
\usepackage{dcolumn}% Align table columns on decimal point
\usepackage{bm}% bold math
\usepackage{hyperref}% add hypertext capabilities
\hypersetup{colorlinks=true,urlcolor=blue,citecolor=blue}
\usepackage{braket} %for bracket notation
\usepackage{soul}

\begin{document}
\title{Retrieval of cavity-generated atomic spin-squeezing after free-space release} 

\author{Yunfan Wu}
\affiliation{Department of Applied Physics, Stanford University, Stanford, California 94305, USA}
\author{Rajiv Krishnakumar}
\affiliation{Department of Applied Physics, Stanford University, Stanford, California 94305, USA}
\affiliation{Department of Physics, PMA, California Institute of Technology, California 91125, USA} 
\author{Juli\'{a}n Mart\'{i}nez-Rinc\'{o}n}
\author{Benjamin K. Malia}
\affiliation{Physics Department, Stanford University, Stanford, California 94305, USA}
\author{Onur Hosten}
\affiliation{Institute of Science and Technology Austria}
\author{Mark A. Kasevich}
\email{kasevich@stanford.edu}
\affiliation{Department of Applied Physics, Stanford University, Stanford, California 94305, USA}
\affiliation{Physics Department, Stanford University, Stanford, California 94305, USA}

\date{\today}

\begin{abstract}
The compatibility of cavity-generated spin-squeezed atomic states with atom-interferometric sensors that require freely falling atoms is demonstrated. An ensemble of $500,000$ spin-squeezed atoms in a high-finesse optical cavity with near-uniform atom-cavity coupling is prepared, released into free space, recaptured in the cavity, and probed. Up to $\sim$ 10 dB of metrologically-relevant squeezing is retrieved for 700 microsecond free-fall times, and decaying levels of squeezing are realized for up to 3 millisecond free-fall times.
The degradation of squeezing results from loss of atom-cavity coupling homogeneity between the initial squeezed state generation and final collective state read-out.  A theoretical model is developed to quantify this degradation and this model is experimentally validated.
\end{abstract}
% The abstract should be <600 characters including spaces: It is.

\maketitle
Atomic sensors, including atom interferometers and atomic clocks, typically operate near the quantum projection noise limit permitted by uncorrelated ensembles of atoms \cite{Rosi_2014,Sorrentino_2014,Bloom_2014,Hinkley_1215}. This limit can be overcome using quantum entanglement. For example, with spin-squeezed states \cite{kitagawa_squeezed_1993} it is feasible to surpass the performance of current state-of-the-art sensors as long as typical atom numbers can be preserved and large levels of squeezing can be obtained. Experimentally, spin-squeezing has been demonstrated through a number of methods \cite{hosten_measurement_2016,sewell_magnetic_2012, muessel_scalable_2014, schleier-smith_states_2010,schmied_bell_2016, maussang_enhanced_2010, esteve_squeezing_2008,Appel_PNAS_2009,cox_deterministic_2016}. To date, the best levels of metrologically-relevant squeezing ($\sim$20 dB) have been obtained in systems where cold atoms are trapped and coupled to an optical cavity and the collective state of the atoms is probed through a cavity mode \cite{hosten_measurement_2016, cox_deterministic_2016}. In these systems, up to order of one-million atoms can be utilized, conforming to standards of well engineered sensors.

Implementation of squeezed-state protocols in sensors with freely moving atoms -- devoid of perturbations due to an external confining potential -- require that the initial squeezed states are prepared in a spatially homogeneous way, i.e. that each atomic spin must contribute equally to the collective spin that is being measured. Otherwise, the retrieval of squeezing is hindered: once the atoms are free to move, the information about their individual contributions to the original collective spin is lost, and a different collective spin, which is not necessarily squeezed, is probed \cite{Hu_nonuniform_2015}. Methods to meet the homogeneity requirement in cavity-generated spin squeezing experiments have been studied in Refs. \cite{cox_spatially_2016,hosten_measurement_2016}.

In this Letter we show that by using an optical cavity apparatus specifically designed to enforce homogeneous atom-cavity coupling \cite{hosten_measurement_2016}, squeezing can be generated and retrieved after the atoms are released to millisecond-long duration free falls and recaptured back into the cavity. With this configuration, we experimentally characterize the effect of atom-cavity coupling inhomogeneity on retrieval of spin squeezing.  We develop and experimentally validate a theoretical model which quantifies the degradation of squeezing in terms of experimentally accessible observables.   Prior work has quantified squeezing for inhomogeneously coupled ensembles where the inhomogeneous coupling is fixed for each atom during both the squeezing and retreival measurement sequences \cite{schleier-smith_states_2010,chen_conditional_2011}. In this work, we quantitatively asses the impact of changes in coupling homogeneity between the squeezing and retrieval operations.  This is relevant to a broad class of atomic sensors which seek to exploit squeezing for enhanced noise performance.

\begin{figure}[ht!]
\centering
\includegraphics[width=0.95\linewidth]{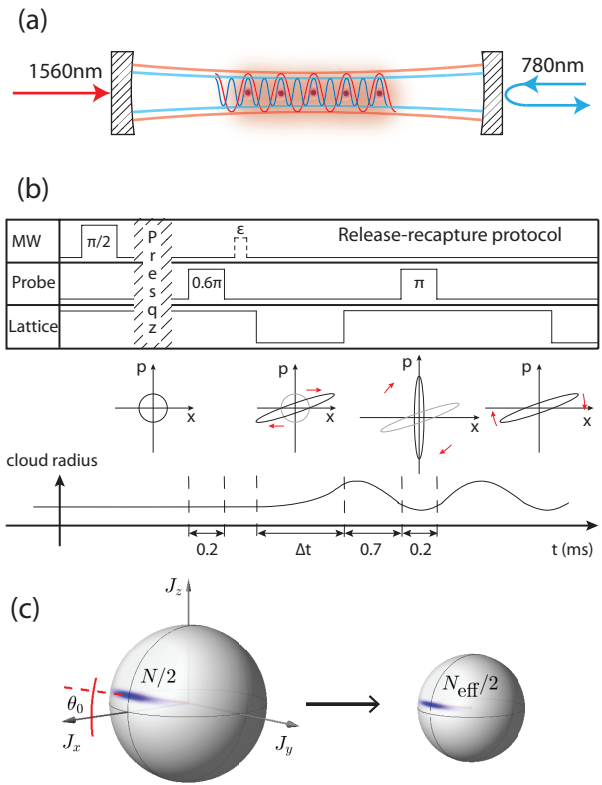}
\caption{\textbf{(a)} Near-uniform atom-cavity coupling of $^{87}$Rb atoms to a 780 nm probe via a commensurate 1560 nm dipole lattice. \textbf{(b)} Release-recapture protocol: timing sequence, illustrations for phase-space evolution of the atomic cloud in the transverse direction of the cavity, and cloud's transverse rms size evolution. In the phase space illustration, clouds are modeled to execute harmonic oscillations when trapped in the lattice. MW: microwaves; Presqz: pre-squeezing; $\pi/2$: composite microwave pulse that prepares the inital superposition states; $0.6\pi$ and $\pi$: probe power expressed in terms of the relative AC-Stark phase shifts induced between the two atomic states; $\Delta t$: free-fall time; $\epsilon$: a small microwave rotation. (\textbf{c}) Inhomogeneous coupling during the readout probe is modelled as reduction of the Bloch sphere radius from $N/2$ to $N_\text{eff}/2$.} 
\label{fig:fig1}
\end{figure}

In our experimental demonstration, we trap $N \sim 500,000$ $^{87}$Rb atoms inside of a 10-cm long high-finesse dual-wavelength cavity that supports both 1560 nm and 780 nm modes. The 1560 nm light forms a 520 $\mu K$ deep 1D optical dipole lattice to trap the 25 $\mu K$ atoms, and the 780 nm light interacts near resonantly with the D$_2$ transition of the atoms to act as a dispersive probe \cite{hosten_measurement_2016, lee_many-atomcavity_2014}. By design, the trapping locations of atoms are aligned with the intensity maxima of the probe standing-wave near the center of the cavity spanning a thousand lattice sites. This alignment gives an almost uniform atom-cavity coupling [\autoref{fig:fig1}(a)]. The magnetically insensitive hyperfine ground states $\ket{F=2, m_F=0}$ and $\ket{F=1, m_F= 0}$ constitute the $\ket{\uparrow}$ and $\ket{\downarrow}$ states for a pseudo-spin 1/2 system. 
 
The atom-cavity detuning is set such that, when interacting with the atoms in the $\left|\uparrow\right>$ state, the cavity resonance acquires a shift equal in magnitude but opposite in sign compared to when it interacts with the $\left|\downarrow\right>$ state atoms. Measurement of this frequency shift by interferometric monitoring of the light reflected from the cavity realizes a quantum nondemolition (QND) measurement of the collective spin $J_z=\sum_{i=1}^N j^{(i)}_{z}$, where $j^{(i)}_{z}=\left(\left|\uparrow\right>_i\left<\uparrow\right|_i-\left|\downarrow\right>_i\left<\downarrow\right|_i\right)/2$ refers to the z-component of the pseudo-spin associated with the $i^{th}$ atom \cite{Kuzmich_1998,noauthor_see_nodate}.  In our work,  a single spin flip between the $\left|\uparrow\right>$ and $\left|\downarrow\right>$ states results in a  $\sim$ 5.6 Hz shift in the cavity resonance frequency, as determined from our cavity parameters and independently verified by measuring the quantum projection noise for coherent spin states \cite{hosten_measurement_2016}.    

For the generation of spin squeezed states, we follow the procedure in \cite{hosten_measurement_2016} with the added step of a free space release-recapture of the atoms between the preparation and readout probes. A one-axis twisting squeezing procedure~\cite{Leroux_PRL_2010,Hosten_science_2016} (or presqueezing) is performed before the first QND measurement (preparation probe) to allow for squeezing of large number of atoms~\cite{hosten_measurement_2016}. After release and recapture, a second QND measurement (readout probe) is performed to observe the collective spin. The preparation probe is chosen to be weaker than the readout probe to obtain large state coherence. In this work, the maximum recovered metrologically-relevant squeezing of this back-to-back measurement protocol is limited to $\sim13$ dB for such a configuration in absence of release. 

The release-recapture protocol is shown in \autoref{fig:fig1}(b). The trapping lattice is turned off after the first probe to release the spin-squeezed atoms in free space. The lattice switching time is 50 $\mu s$, which is adiabatic for the motion in the longitudinal trapping direction but sudden for the transverse one. After a variable free-fall time $\Delta t$ accompanied by a ballistic expansion ($\sim$5 cm/s from a 17 $\mu$m rms radius), the lattice is turned on again to recapture the atoms. After the recapture, the atomic cloud size and position starts oscillating in the transverse direction of the 1D lattice. The maximum recovered squeezing is achieved when the readout probe is turned on while the atomic cloud is maximally compressed during such oscillations.   

The ballistic expansion of the atom cloud and acceleration due to gravity lead to an asymmetry in the atom-cavity coupling between the two probes. This coupling inhomogeneity for the readout probe grows with the free-fall time due to the (fixed) Gaussian spatial profile of the cavity mode. Therefore, the readout probe measures a different observable than the preparation probe that degrades the observed squeezing. In principle one can engineer more advanced release-recapture sequences to better preserve the symmetry in the atomic cloud shape between the two probes. However, such attempts lead to marginal improvements in recovered squeezing (see supplementary material \cite{noauthor_see_nodate}). 
Following the formalism of \cite{Hu_nonuniform_2015}, and as detailed in the supplemental material, the loss of homogeneity can be treated as an effective atom loss irrespective of the details of the coupling inhomogeneity: $N \rightarrow N_\text{eff}$ and $J_z \rightarrow J_{z,\text{eff}}$. 

In earlier squeezed-state measurement-based metrology demonstrations, where the atom-cavity coupling did not substantially change between the state-preparation and readout probes, squeezing efficacy could be directly assessed through comparison of collective spin measurements resulting from back-to-back QND probes. In the current experiment, this protocol is no longer useful, as the effective atom number, and hence the cavity frequency shift, changes between probes.  In order to accommodate this change, we translate observed cavity shifts into Bloch vector angles:  $\theta_0 = J_{\text{z,0}}/(N/2)$ and  $\theta_\text{eff} = J_{z,\text{eff}}/(N_\text{eff}/2)$ for the near homogeneously coupled state preparation and inhomogeneously coupled readout probes respectively (in this experiment $J_{\text{z,0}} \ll N$ and $J_{z,\text{eff}} \ll N_{\text{eff}}$). While the measured cavity shifts differ, the mean values for $\theta_0$ and $\theta_{\text{eff}}$ is preserved between preparation and readout probes (i.e. $\langle \theta_0 \rangle = \langle \theta_\text{eff} \rangle$) \cite{noauthor_see_nodate}.  

\begin{figure}[htb]
\centering
\includegraphics[width=\linewidth]{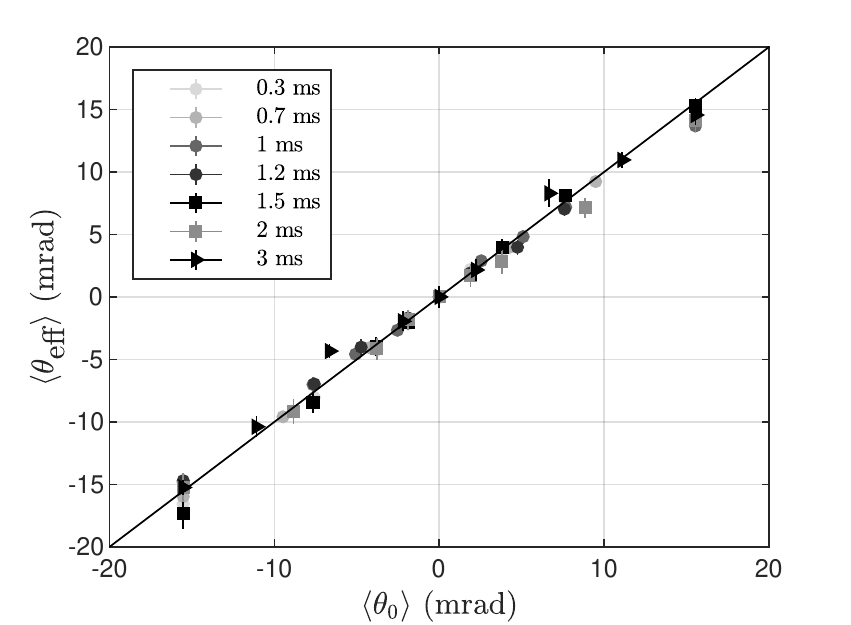}
\caption{Relation between $\langle\theta_\text{eff}\rangle$ and $\langle\theta_0\rangle$. The slope of a linear regression line for each free-fall time (with zero intercept, not shown) agrees with the theoretical value of 1 to within 10\%. Black line: function $\langle\theta_\text{eff}\rangle = \langle \theta_0\rangle$.}
\label{fig:fig0}
\end{figure}

We demonstrate this equivalence experimentally by inserting an additional small microwave rotation about the $y-$axis between the preparation and readout probes  [see \autoref{fig:fig1}(b)] in order to prepare a Bloch vector angle $\theta_0$ away from the equator in the $x-z$ plane [see \autoref{fig:fig1}(c)]. We then compare $\langle\theta_\text{eff}\rangle$ to $\langle \theta_0 \rangle$ at different free-fall times. Since $\theta_0$, $\theta_\text{eff} \ll 1$, these angles can be determined experimentally from the ratio of the observed cavity shift to the maximal shift observed when atoms are prepared in the $\ket{\downarrow}$ state. The result is shown in \autoref{fig:fig0}, which shows that the angle $\langle \theta_{\text{eff}} \rangle$ can be retrieved with better than $\sim 10\%$ error. For this data, we use small atom numbers (50,000 atoms) so that the cavity shifts remain well within the linear response of the homodyne cavity readout.

\begin{figure}[htb]
\centering
\includegraphics[width=\linewidth]{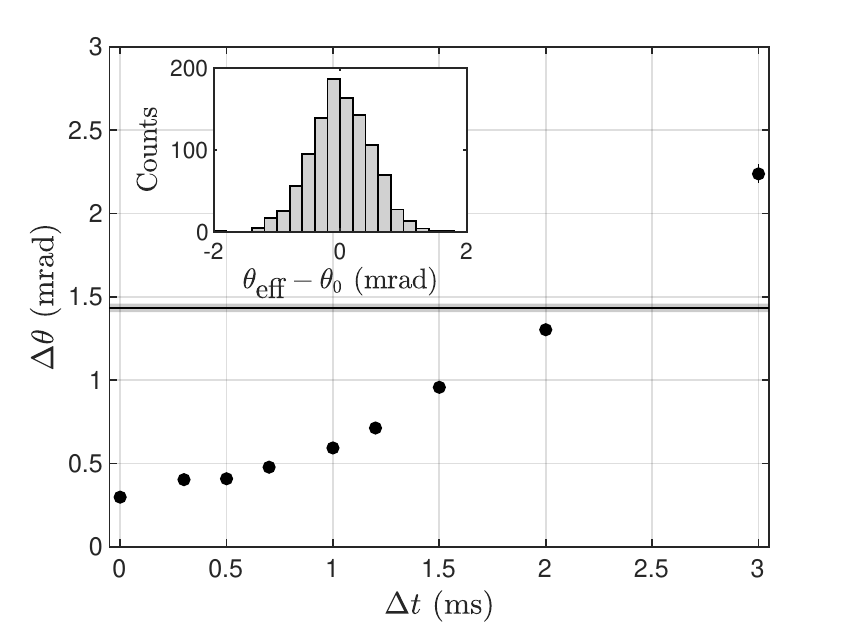}
\caption{Single-shot phase sensitivity $\Delta \theta$ as a function of free-fall time $\Delta t$. $\Delta t=0$ indicates no release. Circles show $\Delta\theta$ for spin squeezed state. The solid line is the quantum projection noise level given by initial atom number $N$. The width of the shaded region is given by the uncertainty in determination of $N$ (68\% confidence interval). Some of the error bars are smaller than the plotted data points. Inset: histogram of $\theta_\text{eff}-\theta_0$ at $\Delta t = 0.7$ms.
}
\label{fig:fig2}
\end{figure}

The phase resolution $\Delta \theta$ of the squeezing protocol is determined experimentally from the measured values of $\theta_\text{eff}$ and $\theta_0$ over an ensemble of measurements.  Specifically, $(\Delta \theta)^2 = \text{var}(\theta_{\text{eff}} - \theta_0)$.  \autoref{fig:fig2} shows the measured resolution $\Delta\theta$ for ensembles of $N\sim 500,000$ spin-squeezed atoms (circles) as a function of free-fall time $\Delta t$ when the state is near the equator of the Bloch sphere. Each data point for $\Delta\theta$ is obtained using more than $700$ independent measurements (inset \autoref{fig:fig2}). The best observed angle sensitivity is $298(8)\,\mu$rad.  As expected, $\Delta \theta$ increases with free-fall time since the inhomogeity increases with this time. For comparison, the projection noise level associated with a coherent spin state with $N$ atoms is also shown in \autoref{fig:fig2}.  $N$ is determined from the observed cavity frequency shift, following Ref. \cite{Hosten_science_2016}. 

Remarkably, the observed loss in phase resolution -- fundamentally associated with the coupling inhomogeneity of the readout probe -- can be accurately described by a model which parameterizes this inhomogeneity in terms of a single parameter,  the effective atom number $N_{\text {eff}}$, regardless of the detailed structure of this inhomogeneity \cite{noauthor_see_nodate}.  Explicitly, $N_{\text{eff}} \equiv {(\sum_{i=1}^N \eta_i)^2}/{\sum_{i=1}^N \eta_i^2}$, where $\eta_i\in[0,1]$ is the fractional coupling of the $i^\text{th}$ atom in the ensemble and $N$ is the total atom number.  From this definition, it can be shown that 
\begin{align}
	(\Delta\theta)^2\simeq   \frac{p_\text{eff}}{ N(1-p_\text{eff})}+\sigma_1^2+\sigma_2^2,
    \label{eq:angle_resolution}
\end{align}
where $p_\text{eff} \equiv 1 - N_\text{eff}/N$ and the terms $\sigma_1^2$ and $\sigma_2^2$ account for the photon shot noise and spin flip noise for the preparation and readout probes respectively \cite{noauthor_see_nodate}. At longer free-fall times, where coupling inhomogeneity plays a significant role, the $\sigma_1$ and $\sigma_2$ terms are dominated by the $p_{\text{eff}}$ term associated with this inhomogeneity. In the limit $p_{\text{eff}} \rightarrow 0$, the expression approaches the noise of the initial spin squeezed state. 

\begin{figure}[htb]
\centering
\includegraphics[width=\linewidth]{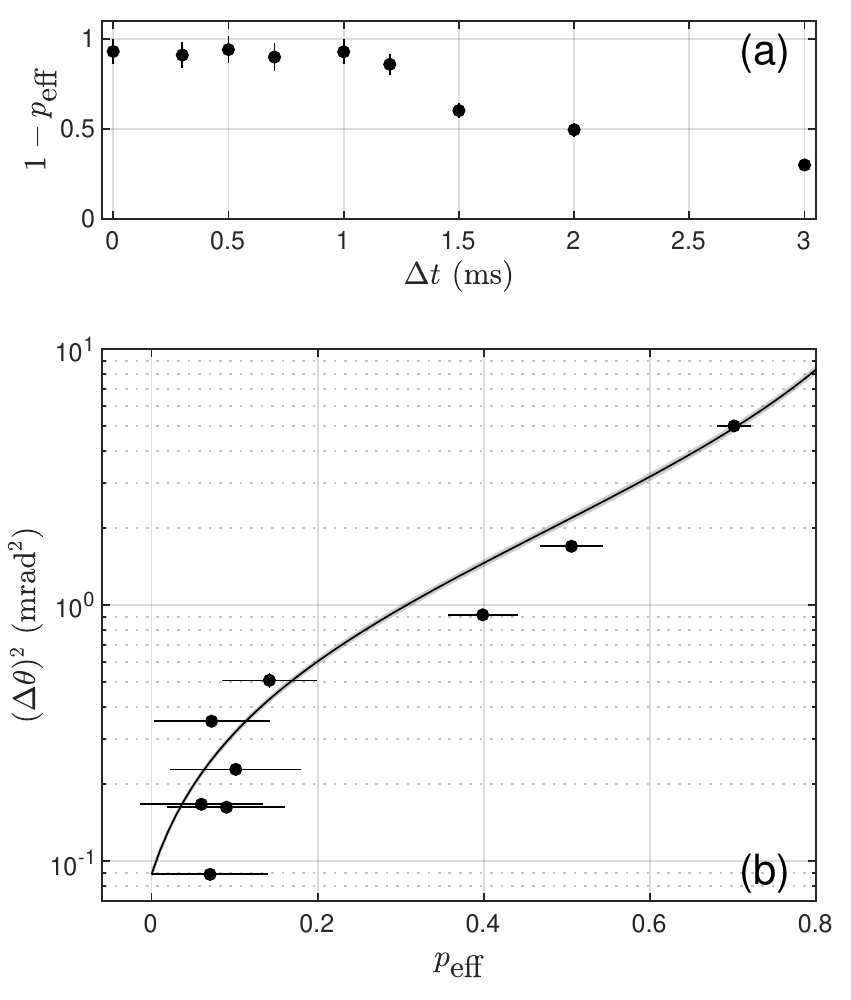}
\caption{(a) Measured value of $1-p_\text{eff}$ as a function of $\Delta t$. (b) $(\Delta \theta)^2$ as a function of effective atom loss probability $p_\text{eff}$. The solid line is a theory curve for $\bar{N}=500,000$ and $\sqrt{\sigma_1^2+\sigma_2^2}=298\,\mu$rad using \autoref{eq:angle_resolution}. The shaded region is due to the uncertainty in determination of $N$ and $\sqrt{\sigma_1^2+\sigma_2^2}$ (width indicates 68\% confidence interval). Error bars show 68\% confidence interval. Some of the error bars in y axis are smaller than the plotted point.}
\label{fig:fig3}
\end{figure}

Comparison of the data shown in \autoref{fig:fig2} with \autoref{eq:angle_resolution} requires experimental determination of $p_{\text {eff}}$.  This can be done by noting that $N_{\text{eff}}$ is defined such that the projection noise in the corresponding effective spin component $J_{z,\text{eff}}$ for an initially prepared coherent spin state is $\text{var}(J_{z,\text{eff}}) = N_{\text{eff}}/4$ while $|J_\text{eff}| = N_\text{eff}/2$ \cite{cox_deterministic_2016}.  Combining these definitions yields $\text{var}(J_{z,\text{eff}})/|J_\text{eff}|^2 = 1/N_{\text{eff}}$, which allows determination of $N_\text{eff}$ through measurement of the ratio $\text{var}(J_{z,\text{eff}})/|J_\text{eff}|^2$ for coherent states. We experimentally determine the value of this ratio through the ratio of the observed fluctuations in the homodyne signal for a coherent spin state [proportional to $\text{var}(J_{z,\text{eff}})^{1/2}$] and the overall cavity shift observed when the atoms are instead prepared in the $\ket{\downarrow}$ state (proportional, with the same constant of proportionality, to $|J_\text{eff}|$).  Combined with an initial measurement of $N$, $p_{\text{eff}}$ is thus determined.  \autoref{fig:fig3}(a) shows the resulting inferred values of $p_{\text{eff}}$ as a function of free-fall time.  Since $p_\text{eff}$ is independent of initial atom number, we determine $p_{\text {eff}}$ with smaller ensembles of $N \sim 100,000$ atoms to avoid the influence of microwave rotation noise in the preparation of the initial coherent spin state.    Substitution of the measurement of $p_\text{eff}$, together with the known value of $\sigma_1^2+\sigma_2^2$, into \autoref{eq:angle_resolution} leads to  \autoref{fig:fig3}(b). 

\begin{figure}[htb]
\centering
\includegraphics[width=\linewidth]{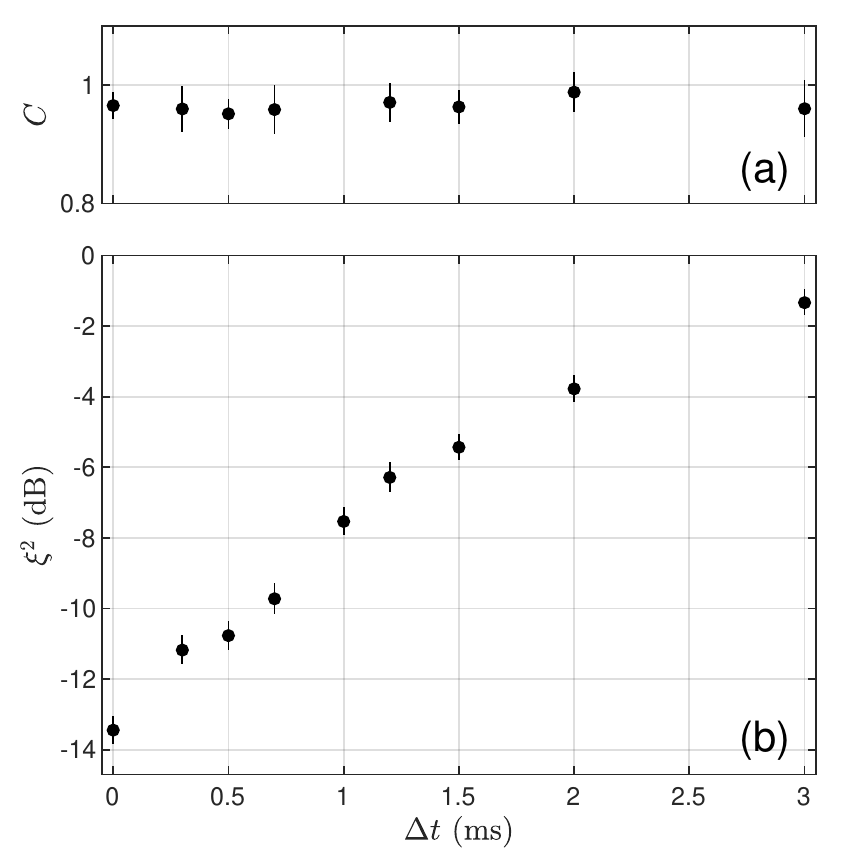}
\caption{(a) Coherence of atoms during free falling as a function of free-fall times $\Delta t$ and (b) Retrieved metrologically-relevant squeezing $\xi^2$. Error bars indicates 68\% confidence interval.
}
\label{fig:fig4}
\end{figure}

Metrologically-relevant squeezing can be quantified with the Wineland parameter $\xi^2$ ~\cite{wineland_spin_1992,schleier-smith_states_2010}, which compares the angle resolution $\Delta \theta$ obtained with a squeezed state to that obtained with an (unsqueezed) coherent spin state having the same number of atoms and also accounts for coherence $C$ of the ensemble.  Using the notation defined above, the Wineland parameter $\xi^2$ takes the form
\begin{eqnarray}\label{xi}
        \xi^2= \bigg ( \frac{\Delta \theta}{1/\sqrt{N_\text{eff}}}\; \frac{1}{C} \bigg )^2.\
\end{eqnarray}
 Experimentally, we characterize $C$ with an additional microwave $\pi/2$ rotation just before recapturing the atoms.  We find $C \simeq 0.96$, independent of free-fall time [see \autoref{fig:fig4}(a)].
 \autoref{fig:fig4}(b) shows the inferred Wineland parameter as a function of $\Delta t$, given the measurements of $\Delta \theta$ and $N_\text{eff}$ with $N\sim500,000$ initial atoms. Despite substantial loss of homogeneity, metrologically relevant squeezing is observed to persist for time scales as long as 3 ms. We can recover most of the initial squeezing %(up to $9.8^{+0.5}_{-0.4}$ dB) 
for shorter free-fall times ($\Delta t < 1$ ms).

Although the successful retrieval of squeezing was limited to $\sim$ 3 ms free-fall times in this work, a substantial extension of this duration should be possible using a significantly colder atomic ensemble and an optimized cavity orientation with respect to gravity. We expect the model used to quantify the loss of squeezing to be useful in designing and predicting the performance of future squeezed-state sensors. 

%Nocites here to add extra references from SM to the bibliography.
%\nocite{vitagliano2016}
%\nocite{dellantonio}
\begin{acknowledgments}
Acknowledgements. We thank Nils Engelsen for comments on the manuscript. This work was supported by the Office of Naval Research, Vannevar Bush
Faculty Fellowship Department of Energy, and Defense Threat Reduction Agency. R.K. was partly supported by the AQT/INQNET program at Caltech.
\end{acknowledgments}
%\bibliography{FreeSpacePaper}

%

\clearpage
\newpage
\onecolumngrid 
\section{supplementary material}

\subsection{Modeling the Effects of Inhomogeneous Atom-cavity Coupling}
\label{sec:inhomogeneity}
Following the formalism described in \cite{Hu_nonuniform_2015}, we introduce effective observable $J_\text{z,eff}$ and effective atom number $N_\text{eff}$  to model the effects of inhomogeneity so that the standard relations $\langle J^{ \text{max}}_{z,\text{eff}}\rangle=N_{\text{eff}}/2$ and $\text{var}(J_{z,\text{eff}})=N_{\text{eff}}/4$ are satisfied. Here, $J^{\text{max}}_{z,\text{eff}}$ is shorthand for $J_{z,\text{eff}}$ when the quantum state gives $j_z^{(i)}=1/2$ for all $i$. This state is represented by the Bloch vector that is pointing to the north pole of the Bloch sphere. This definition is consistent with that of identically prepared uncorrelated particles where $\langle J_z^\text{max}\rangle = N/2$ and $\text{var}(J_z) = N/4$. Accordingly, we write $J_{z,\text{eff}}= \langle\eta\rangle_e\sum_i\eta_ij_z^{(i)}/\langle\eta^2\rangle_e=N_\text{eff}\sum_i\eta_ij_z^{(i)}/N\langle\eta\rangle_e$ and $N_{\text{eff}} = N\langle\eta\rangle_e^2/\langle\eta^2\rangle_e
$ where $\eta_i\in[0,1]$ is the fractional coupling of the $i^\text{th}$ atom and $N$ is the actual total atom number. Here $\langle \cdot \rangle_e$ is the ensemble average of a quantity over the atoms, e.g $\langle\eta\rangle_e=\sum_i\eta_i/N$. Given the total cavity frequency shift $\delta_{\text{cav}}=\sum_i\delta_0\eta_ij_z^{(i)}=\delta_{\text{eff}}J_{z,\text{eff}}$, a shift per spin flip can be defined as $\delta_{\text{eff}}=\delta_0\langle\eta^2\rangle_e/\langle\eta\rangle_e$.

Now we relate the observations of $J_{z,\text{eff}}$ to those of $J_z$. We rely on the fact that $\langle j_z^{(i)}\rangle$ and $\langle j_z^{(i)}j_z^{(i\neq k)}\rangle$ are independent of $i$ and $k$ owing to the homogeneity of the prepared states. Thus, we get
\begin{eqnarray}
    \langle J_z\rangle&=&\sum_i\langle j_z^{(i)}\rangle=N\langle j_z^{(i)}\rangle\\
    \langle J_z^2\rangle&=&\sum_{i,k}\langle j_z^{(i)}j_z^{(k)}\rangle=\sum_{i\neq k}\langle j_z^{(i)}j_z^{(k)}\rangle +\sum_i\langle j_z^{(i)2}\rangle=N(N-1)\langle j_z^{(i)}j_z^{(i\neq k)}\rangle+N\langle j_z^{(i)2}\rangle\\
    \langle J_{z,\text{eff}}\rangle&=&\frac{\langle\eta\rangle_e}{\langle\eta^2\rangle_e}\sum_i\eta_i\langle j_z^{(i)}\rangle=\frac{N_\text{eff}}{N}\langle J_z\rangle\\
    \langle J_{z,\text{eff}}^2\rangle&=&\frac{\langle\eta\rangle_e^2}{\langle\eta^2\rangle_e^2}\sum_{i,k}\eta_i\eta_k\langle j_z^{(i)}j_z^{(k)}\rangle=\frac{N_\text{eff}}{N}\langle J_z^2\rangle-\frac{N_\text{eff}}{N}(1-\frac{N_\text{eff}}{N})\langle j_z^{(i)}j_z^{(i\neq k)}\rangle N^2\\
    \langle J_{z,\text{eff}}J_z\rangle &=& \frac{N_\text{eff}}{N}\langle J_z^2\rangle
\end{eqnarray}

Defining Bloch vector angles $\theta_0=J_z/(N/2)$ and $\theta_\text{eff}=J_\text{eff}/(N_\text{eff}/2)$ for the homogeneous and inhomogeneous probe respectively, we prove that $\langle\theta_\text{eff}\rangle=\langle J_{z,\text{eff}}\rangle/(N_\text{eff}/2)=\langle J_z\rangle/(N/2)=\langle\theta\rangle$. %This justifies that a mean rotation of the atomic state between the probe and readout pulses can be represented by $\langle\theta_{\text{eff},2}\rangle - \langle\theta_{0,1}\rangle$. 

Assuming the atomic state is lying close to the equator of the Bloch sphere and each location in space contains only one atom, we obtain $\langle j_z^{(i)}\rangle=\sigma$ ($\sigma\ll 1$) and $\langle j_z^{(i)2}\rangle=1/4+\sigma^2$. Consequently, $\langle j_z^{(i)}j_z^{(i\neq k)}\rangle=(\xi^2-1)/4(N-1)+\sigma^2$, where the variance of $J_z$ with respect to the coherent spin state (CSS) noise is defined as $\text{var}(J_z)/(N/4)=\xi^2$. In addition, $\langle j_z^{(i)2}\rangle-\langle j_z^{(i)}j_z^{(i\neq k)}\rangle= [1+(1-\xi^2)/(N-1)]/4\approx 1/4$, where the approximation is valid if the $J_z$ noise is close to the CSS noise; for example for an $J_z$ noise level 20dB above CSS noise level ($\xi^2=100$), the fractional correction is only $2\times 10^{-4}$ for $500\,000$ atoms. Therefore, 
\begin{eqnarray}
    \text{var}(J_{z,\text{eff}})&=&\frac{N_\text{eff}^2}{N^2}\text{var}(J_z)+N_\text{eff}(1-\frac{N_\text{eff}}{N})(\langle j_z^{(i)2}\rangle-\langle j_z^{(i)}j_z^{(i\neq k)}\rangle)\\
    &\approx&(1-p_\text{eff})^2\text{var}(J_z)+p_\text{eff}\frac{N_\text{eff}}{4}\label{eqn:var_Jz_eff}\\
    \text{var}(J_{z,\text{eff}}-J_z) &\approx& p_\text{eff}^2\text{var}(J_z)+p_\text{eff}\frac{N_\text{eff}}{4} 
\end{eqnarray}

Here we define an effective atom loss probability $p_\text{eff}$ such that the remaining atom number is $N_\text{eff}=(1-p_\text{eff})N$. %We find that $p_\text{eff}$ due to atom-cavity coupling inhomogeneity is effectively an atom loss. 

\subsection{Modeling Back-to-back Conditional Measurement}
\label{sec:back-to-back}
Since $J_z$ is inferred from the X-quadrature of a probe field through a calibrated discriminator, we label the X-quadrature of the two probes in the back-to-back measurements as $X_1$ and $X_2$. Since these two field modes are uncorrelated, $\text{cov}(X_1,X_2)=0$. We also label the collective spin operators for the atoms during the first and second measurements as $J_z$ and $J_\text{eff}$ respectively. Since before sending probes to the atoms, there are no correlations between the spins and the fields, i.e., $\text{cov}(J_{z,\text{eff}},X_1)=\text{cov}(J_z,X_1)=\text{cov}(J_{z,\text{eff}},X_2)=\text{cov}(J_z,X_2)=0$;  after probe interactions, the quadrature operators (in the Heisenberg picture) become 
%\begin{equation}
%    \begin{align}
    \begin{eqnarray}
        X_1\rightarrow X_1'&=&D_1J_z+X_1\\
        X_2\rightarrow X_2'&=&D_2J_{z,\text{eff}}+X_2
    \end{eqnarray}
%    \end{align}
%\end{equation}
where $D_i$ are calibrated discriminators determined by the strength of the probes. Thus the inferred $J_{z}$ values from the two probes are $J_{z1}=X_1'/D_1$ and $J_{z2}=X_2'/D_2$. The variance of the inferred $J_z$ difference is 
%\begin{equation}
%    \begin{align}
\begin{eqnarray}
    \text{var}(J_{z2}-J_{z1})&=&\text{var}(J_{z,\text{eff}}-J_z)+\frac{1}{D_1^2}\text{var}(X_1)+\frac{1}{D_2^2}\text{var}(X_2)\\
    &\approx&p_\text{eff}\frac{N_\text{eff}}{4}+p_\text{eff}^2\text{var}(J_z)+\sigma_{X1}^2+\sigma_{X2}^2
%    \end{align}
%\end{equation}
\end{eqnarray}
Here $\sigma_{X1}^2$ and $\sigma_{X2}^2$ are the squares of $J_z$ resolutions by the first and the second probes respectively due to optical shot noise and spin flips. The second term is an excess noise due to changes in $J_z$ between the two probes whose value depends on the outcome of the first probe which itself is random. 

To eliminate this noise, we work with Bloch vector angle $\theta$. The angles inferred from the Jz measurements are thus defined as $\theta_1=J_{z1}/(N/2)=2X_1'/ND_1$ and $\theta_2=J_{z2}/(N_\text{eff}/2)=2X_2'/N_\text{eff}D_2$ respectively. The variance of the difference between the two inferred angles is
\begin{eqnarray}
    \text{var}(\theta_2-\theta_1)&=&\text{var}(\theta_\text{eff}-\theta_0)+\frac{1}{D_1^2 N^2/4}\text{var}(X_1)+\frac{1}{D_2^2 N_\text{eff}^2/4}\text{var}(X_2)\\
&\approx&\frac{p_\text{eff}}{N_\text{eff}}+\sigma_{\theta1}^2+\sigma_{\theta2}^2\\
&=&\frac{p_\text{eff}}{N(1-p_\text{eff})}+\sigma_{\theta1}^2+\sigma_{\theta2}^2\label{eqn:main}
\end{eqnarray}

This equation gives the angle resolution for our setup that accounts for the noise due to changes in the atom-cavity coupling between the two probes [$p_\text{eff}/(N(1-p_\text{eff}))$] and noise due to the initial squeezing ($\sigma_{\theta1}^2+\sigma_{\theta2}^2$). \autoref{eqn:main} is what we present in the main paper as Eq. (1) where $\text{var}(\theta_2-\theta_1)$ is replaced by $(\Delta\theta)^2$ and $\sigma_{\theta1}^2+\sigma_{\theta2}^2$ is replaced by $\sigma_1^2+\sigma_2^2$.

\subsection{Cavity Readout}
We measure resonance frequency shift of a high finesse cavity to infer the atomic state and calibrate for the atom number in this experiment. 
\subsubsection{Linewidth broadening}
\label{subsec:linewidth}
The cavity shift is measured by comparing the time-dependent homodyne signal to a normalized template taken in absence of atoms. Corrections to the nonlinearity of the cavity frequency response and linewidth broadening factor $\kappa_s$ due to scattering by atoms are applied in the same way as in Refs. \cite{nils_thesis_2016,hosten_measurement_2016}. However, the atom-cavity coupling is less homogeneous for the second probe after release-recapture, this effect changes the linewidth broadening factor to $\langle\eta\rangle_e\kappa_s$.

\subsubsection{Measurement of maximum cavity frequency shift and $\langle\eta\rangle_e$}
\label{subsec:eta}
To measure maximum cavity frequency shift and $\langle\eta\rangle_e$, we prepare all the atoms in $\ket{\downarrow}$ state. Since $100\,000$ $\ket{\downarrow}$ state atoms give a cavity resonant frequency shift of $\sim250$KHz, which is far more than the cavity linewidth $\sim10$KHz, we use a different method to measure cavity frequency shift than the back-to-back method used in the main paper \cite{hosten_measurement_2016}. In this method, the 780 nm probe is on continuously through the release-recapture (RR) sequence and its frequency is scanned from $50$KHz to $-50$KHz off resonant in $200\mu s$ that overlaps with the time when the second probe is on during a RR sequence in the back-to-back method. This scan gives a dispersive signal by the homodyne detection when no atom is loaded. This dispersive signal changes its shape due to the added $\ket{\downarrow}$ atoms. Adjusting the starting frequency of the frequency scan brings the dispersive signal back to empty cavity shape (the frequency range of the scan remains $100$KHz). The amount of adjusted frequency tells the maximum cavity frequency shift caused by the amount of added atoms which is readout by fluorescent imaging. 

Due to the bandwidth of the laser frequency lock, the maximum adjustment on the starting frequency is limited to $\sim900$KHz. This also limits the maximum added atom number to $\sim300\,000$. To get the maximum cavity frequency shift for all different atom number ($\sim50\,000$, $100\,000$ and $500\,000$ in this work) and reduce uncertainties, we measure this shift as a function of number of added atoms and fit this function with a straight line (as expected). The maximum cavity frequency shift that corresponds to a specific atom number is found by interpolation or extrapolation. 

We measure and linearly fit this frequency shift as a function of atom number for each free-fall time. The slope of each fitting is proportional to $\langle\eta\rangle_e$ with the same constant of proportionality. This constant can be calculated by fitting of the zero free-fall time data whose $\langle\eta\rangle_e = 0.9254$ is known \cite{Hosten_science_2016}. $\langle\eta\rangle_e$ for other free-fall times are thus calculated. 

\subsection{Coherence Measurement}
The coherence of the atomic state is measured by a microwave-induced Ramsey oscillations. The first Ramsey $\pi/2$ pulse is the composite $\pi/2$ microwave pulse at the beginning of the experimental sequence where the atoms are trapped by the lattice. The second $\pi/2$ pulse is applied after atoms free falling for a certain amount of time. The phase of this pulse is adjusted to be roughly $90^\circ$ to the atoms by a frequency shift key and is scanned by a small amount. This scan covers the bottom (top) part of the Ramsey fringes. A quadratic fit is applied to this part of the fringes. The highest point of this fit stands for the coherence. The prediction interval of that point gives the uncertainty of the coherence measurement. Every step in the sequence before the release remains the same as the squeezing measurement.  Since the cavity resonance is shifted out of the cavity linewidth, a fluorescence population  spectroscopy is used to measure the collective $J_z$ \cite{Hosten_science_2016}. This coherence is measured with $\sim500\,000$ atoms. 

\subsection{Delta-Kick Protocol}
\begin{figure}[htb]
\centering
\includegraphics[width=12cm]{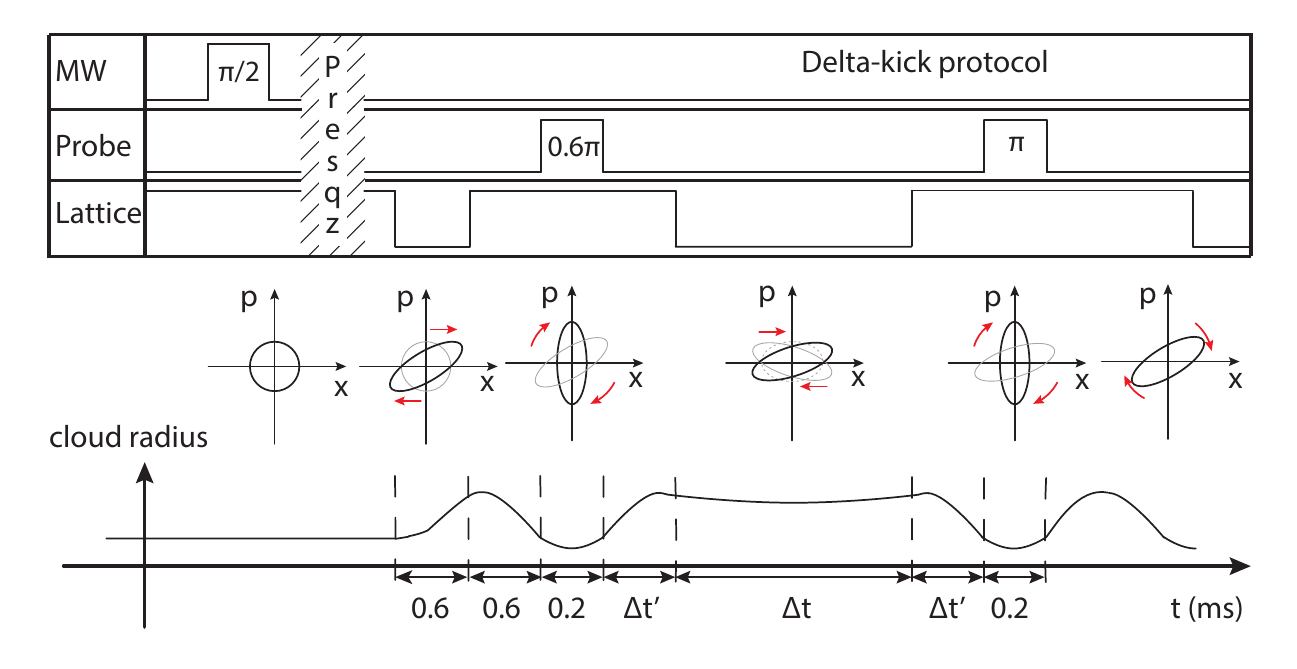}
\caption{\textbf{(a)} Delta-kick protocol: timing sequence; illustration for phase-space evolution of the atomic cloud in the transverse direction of the cavity; illustration for cloud's rms transverse size evolution. In the phase space illustration, clouds are modeled to execute harmonic oscillations when trapped in the lattice. MW: microwaves; Presqz: pre-squeezing; $\pi/2$: composite microwave pulse that prepares the 50-50 superposition states; $0.6\pi$ and $\pi$: probe power expressed in terms of the AC-Stark phase shifts induced between the two atomic states; $\Delta t'$: short free-fall time for atomic phase space reshaping; $\Delta t$: free-fall time. The choices of $\Delta t'$ and $\Delta t$ are shown in \autoref{table:dk}.}
\label{fig:fig_supp}
\end{figure}

In the `delta-kick' protocol (\autoref{fig:fig_supp}), we utilize a series of lattice on-off sequences to reshape the phase space distribution of the cloud. The timing in this protocol is based on a simulation assuming the lattice is a harmonic trap and experimentally chosen. See \autoref{table:dk}. This protocol gives less coherence ($<92\%$) compared to the RR protocol irrespective of free-fall time, because the extra lattice on-off sequences induce more inhomogeneous AC-Stark shifts on the atoms. The metrologically-relevant squeezing decreases to $0$ in less than 5ms free-fall time.

\begin{center}
\begin{table}
	\begin{tabular}{|l|l|l|l|l|}
    \hline
     $\Delta t'$ (ms) &0.8&0.7&0.6&0.5\\\hline
     $\Delta t$ (ms) &0.4&0.7&1.2&1.4, 2, 3, 4, 5, 6, 7\\\hline
	\end{tabular}
	\caption{The delta-kick protocol timing sequence. $\Delta t^\prime$ stands for the short free-fall time before the measurement sequence to reshape the phase space distribution of the cloud. $\Delta t$ is the free-fall time that is used for freespace atomic sensors.}
\label{table:dk}
\end{table}
\end{center}

\subsection{Uncertainty Calculations}

We calculate uncertainties using standard error propagation formula where the uncertainty of each quantity stands for 0.68 confidence interval. These related quantities are assumed to be independent and their errors are assumed to follow normal distributions. In determining the uncertainties of $\Delta\theta_\text{eff}$ and squeezing, only statistics uncertainty on cavity frequency measurements is considered. This uncertainty is estimated by a bootstrapping method where we re-sample $10\,000$ times from the measured frequency distribution, calculate $10\,000$ standard deviations from the resampled data and estimate this uncertainty as the standard deviation of these $10\,000$ standard deviations.

\end{document}